\documentclass[twocolumn,showpacs,preprintnumbers,amsmath,amssymb,pre,
superscriptaddress]{revtex4}

\usepackage{graphicx}
\usepackage{dcolumn}
\usepackage{bm}
\usepackage{psfrag}

\usepackage{verbatim}

\begin{document}

\title{Phase behavior of a nematic liquid crystal in contact with a chemically 
and geometrically structured substrate}

\author{L. Harnau} \author{S. Kondrat}
\affiliation{Max-Planck-Institut f\"ur Metallforschung,  
Heisenbergstr.~3, D-70569 Stuttgart, Germany}
\affiliation{Institut f\"ur Theoretische und Angewandte Physik,               
Universit\"at Stuttgart, Pfaffenwaldring 57, D-70569 Stuttgart, Germany}

\author{A. Poniewierski}
\affiliation{Institute of Physical Chemistry, Polish Academy of
Sciences, Kasprzaka 44/52, 01-224 Warsaw, Poland}

\date{\today}

\begin{abstract}

A nematic liquid crystal in contact with a  grating surface possessing
an alternating stripe pattern of locally homeotropic and planar anchoring is studied 
within the Frank--Oseen model. The combination of both chemical and geometrical
surface pattern leads to rich phase diagrams, involving a homeotropic, a planar, and
a tilted nematic texture. The effect of the groove depth and the anchoring strengths 
on the location and the order of phase transitions between different nematic 
textures is studied. A zenithally bistable nematic device is investigated by confining 
a nematic liquid crystal between the patterned grating surface and a flat substrate 
with strong homeotropic anchoring.

\end{abstract}

\pacs{64.70.Md, 61.30.Cz}

\maketitle

\section{Introduction}

The contact of nematic liquid crystals (NLCs) with structured solid
substrates  offers the possibility to manipulate the orientation of the
adjacent  liquid crystal molecules in a controlled way. For example,
substrates with  topographic structures influence anchoring of liquid
crystal molecules and induce elastic distortions and flexoelectric
polarizations within a  contacting NLC. Alignment of liquid crystal
molecules can also be governed by  the pattern of boundary lines between
different regions on a flat substrate and the NLC elasticity. Adjacent
regions  of planar and homeotropic anchoring within a single substrate
have been  created, for example, by micro-contact printed polar and apolar thiols,
respectively, on  an obliquely evaporated ultra-thin gold layer. Whereas
many experimental and  theoretical studies have focussed on the
understanding of the interactions  and phase behavior of NLCs in contact
with {\it either} geometrically structured  substrates (see e.g.,
\cite{berr:72,faet:87,bari:96,moce:99,four:99,barb:99,brow:00,patr:02,
chun:02,lu:03,parr:03,brow:04,parr:04}) {\it or} chemically patterned
substrates (see e.g., \cite{ong:85, barb:92, qian:96, qian:97, gupt:97,
lee:01,kond:01,wild:02, wild:03, kond:03, park:03, zhan:03, tsui:04,
yu:04,poni:04}), NLCs near geometrically structured {\it and} chemically
patterned substrates have  not been investigated yet. In view of the
rich behavior of NLCs even on   geometrically structured or chemically
patterned substrates, a combination  of both surface treatments may open
new possibilities for an improved performance  of NLC cells. Here we
study a NLC in contact with a grating  surface possessing an alternating
stripe pattern of locally homeotropic and planar anchoring within the
Frank--Oseen model \cite{fran:58,dege:93},  whereby the anchoring energy
function is given by the Rapini--Papoular expression \cite{rapi:69}.
Taking into account both chemical and geometrical surface pattern is
particularly interesting because of the possibility of bistable 
anchoring, which is important for low power consumption liquid crystal
displays. In bistable nematic devices there are two stable nematic
director orientations which have  substantially different tilt angles.
The zenithally bistable nematic devices that  have been studied recently
consist of a NLC confined between a grating surface  and a flat surface
\cite{parr:03,brow:04,parr:04,davi:02}. These surfaces  are coated with
a homeotropic agent. While such a device is characterized by both  a
high shock stability and a low power consumption, the deep grating
structure  leads to a reduction of the contrast of the display
\cite{krie:02}. On the basis  of our calculations we expect that an
additional chemical surface pattern  allows one to use rather a shallow
grating surface instead of a deep one. 

The structure of the paper is as follows. In Section
\ref{sec:semi-infinite} two models of a patterned grating surface are
introduced and phase diagrams of a NLC in contact with the model
surfaces are discussed. In particular we show how phase transitions 
between different nematic textures vary with the groove depth and the
anchoring strengths. The phase behavior and director profile of a NLC
confined between a patterned grating surface and a flat surface subject
to strong homeotropic anchoring is investigated in
Sec.~\ref{sec:finite}. Our results are summarized in
Sec.~\ref{sec:summary}.

\section{NLC in contact with a chemically and geometrically patterned substrate}
\label{sec:semi-infinite}

We first consider a semi-infinite system consisting of a NLC in contact 
with a single, patterned grating surface. Two models of the surface
grating are investigated. In model ${\mathcal{A}}$, a sinusoidal grating
surface is assumed. This leads to the Euler--Lagrange equation for a
two-dimensional director field, which is then solved numerically.   In
model ${\mathcal{B}}$, we choose a special form of the surface grating 
which, albeit not given explicitly by a simple formula, has the great 
advantage that the problem can be reformulated in terms of a
semi-infinite  system with a flat patterned substrate, and for the
latter we have developed  previously an efficient method of finding the
equilibrium director field \cite{kond:01}.

\begin{figure}[t!]
\begin{center}
\includegraphics[width=8cm]{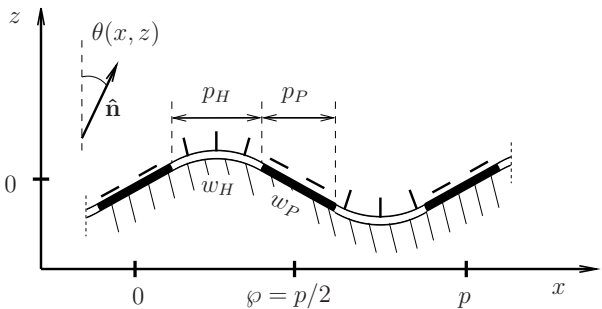}
\end{center}
\caption{The system under consideration consists of a nematic liquid crystal (NLC) 
in contact with a grating surface with an alternating stripe pattern  of
locally homeotropic anchoring (white bars) and homogeneous planar
anchoring  (black bars). The stripes are orientated along the  $y$ axis
perpendicular to the  plane of the figure. The projection of the widths
of the stripes onto the $x$-axis and the anchoring strengths are
designated as $p_H=p/4$, $p_P=p/4$ and $w_H$, $w_P$, respectively, and
the midlines of the planar anchoring stripes are located at $x=np/2$
with $n=0,\pm1,\pm2, \cdots$; $\wp$ is the period of the chemical
pattern which is half the period of the surface grating $p$. The tilt of
the nematic director with respect to the $z$-axis is denoted by
$\theta(x,z)$.}
\label{fig1}
\end{figure}

\subsection{Model ${\mathcal{A}}$: Sinusoidal grating surface}
\label{sub:si:model_A}

Let us consider a NLC in contact with a single grating surface and
assume that the surface profile $z_0(x)$ is given by
\begin{equation} \label{eq1}
z_0(x)=A\sin \left(q x\right)\,,
\end{equation} 
where $A$ is the groove depth and $p=2\pi/q$ is the period. Such a
surface  structure can be fabricated, for instance, using an
interferometric exposure  technique (see Ref.~\cite{brow:04} and
references therein). As Figure \ref{fig1}  illustrates, the surface
exhibits a pattern consisting of alternating stripes  of locally
homeotropic and homogeneous planar anchoring. The system is 
translationally invariant in the $y$ direction. We focus on shallow
grooves for which the nematic director ${\bf\hat n}$ can be constrained
to the plane perpendicular to the direction of the surface grooves. The
orientation of ${\bf\hat n}$ is therefore given solely by the polar
angle $\theta(x,z)$ [see Fig.~\ref{fig1}]. The free energy functional of
the NLC reads  
\begin{eqnarray} \label{eq2}
&&F[\theta]=F_d[\theta]+F_s[\theta_0] \nonumber\\[3pt]
&&=\frac{K}{2}\int^p_0 dx\,\int^\infty_{z_0(x)} dz\,
    [\nabla \theta(x,z)]^2 \nonumber
\\&&+\frac{1}{2}\int^p_0 dx\,
w(x)\frac{[-\sin(\theta_0(x))z'_0(x)+
\cos(\theta_0(x))]^2}{\sqrt{1+(z'_0(x))^2}} 
\end{eqnarray} 
where $\theta_0(x)=\theta(x,z_0(x))$.
The first term on the right hand side of Eq.~(\ref{eq2}) is the distortion 
free energy ($F_d$) \cite{fran:58,dege:93} within the one-elastic-constant 
approximation, and the second term is the surface free energy ($F_s$)
adopting the Rapini--Papoular form 
\cite{rapi:69}. The anchoring strength is specified by a periodic step
function: $w(x)=-w_H$ and  $w(x)=w_P$ for values of $x$ on the
homeotropic  and planar anchoring stripes, respectively (see
Fig.~\ref{fig1}). Equation (\ref{eq2}) completely specifies the free
energy functional for the system under consideration. Minimization of
$F[\theta]$ with respect to $\theta(x,z)$ leads to the Laplace equation
with the following boundary condition at $z=z_0(x)$: 
\begin{eqnarray} \label{eq2a}
&&2K\lim_{z\to z_0(x)}\left[-z'_0(x)\partial_x+\partial_z\right]
\theta(x,z)=\nonumber
\\&&
-\frac{w(x)}{\sqrt{1+(z'_0(x))^2}}
\left\{ [1-(z'_0(x))^2]\sin (2\theta_0(x))\right.\nonumber
\\&&+\left.
2z'_0(x)\cos (2\theta_0(x))\right\}\,,
\end{eqnarray}
and the second boundary condition: $\lim_{z\to \infty} \partial_z \theta(x,z)=0$.
We solve this equation numerically on a sufficiently fine two-dimensional 
$(x,z)$ grid. 

The periodic surface induces a certain anchoring direction (i.e., the
orientation of ${\bf\hat n}$ far from the surface) which we call the
{\it effective anchoring direction} $\theta_a^{(eff)}$, to distinguish
it from the local anchoring directions of different regions   forming
the surface pattern.  In the absence of external fields or competing
surfaces ${\bf\hat n}$ adopts the orientation $\theta_a^{(eff)}$ when
the distance from the  surface is large compared to the periodicity of
the surface pattern. As we know \cite{kond:01}, in the case of a flat
surface ($z'_0(x)=0$) uniform nematic textures can exist: homeotropic
($\theta(x,z)=0$) and planar ($\theta(x,z)=\pi/2$),  whereas it is clear
from Eq.~(\ref{eq2a}) that no uniform texture  ($\theta(x,z)=const$) is
possible if $z'_0(x)\neq 0$. Nevertheless, the homeotropic (H) texture,
defined by $\theta_a^{(eff)}=0$, or planar (P) texture, defined by
$\theta_a^{(eff)}=\pi/2$, can still exist even though $\theta(x,z)\neq
const$ close to the surface. In other words, the presence of surface
grating does not necessarily imply a tilted (T) nematic texture,
corresponding to $0<\theta_a^{(eff)}<\pi/2$. The existence of solutions
corresponding to the H and P textures for arbitrary groove depth follows
from the assumed symmetry of the surface profile and anchoring strength
function: $z_0(x)=z_0(p/2-x)$ and $w(x)=w(p/2-x)$. Since
Eq.~(\ref{eq2a}) is invariant with respect to the transformation $x\to
p/2-x$ if $\theta(p/2-x,z)= n\pi -\theta(x,z)$, where $n=0$ or $n=1$, we
have $\theta_a^{(eff)} = 0$ for the antisymmetric solution ($n=0$) and
$\theta_a^{(eff)} = \pi/2$, for $n=1$. Note that the above argumentation
holds only if the functions $z_0(x)$ and $w(x)$ exhibit the same
symmetry, although even in this case the tilted texture can be more
stable than the H texture or P texture. If $z_0(x)$ and $w(x)$ are not
in phase only the tilted texture is possible.

\subsection{Model ${\mathcal{B}}$: Special form of surface grating}
\label{sub:si:model_B}

\begin{figure}[t!]
\vspace*{-1.2cm}
\begin{center}
\includegraphics[width=8cm]{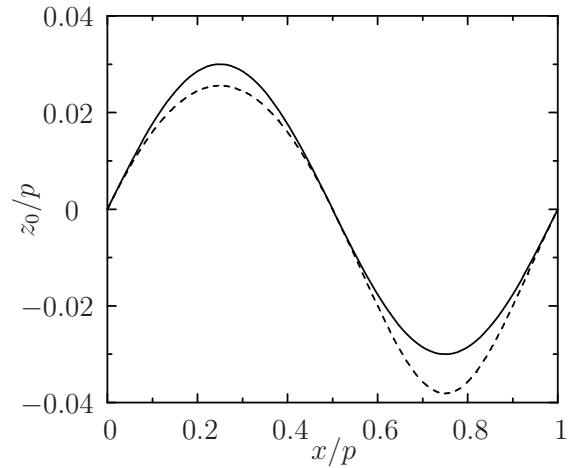}
\end{center}
\vspace*{-4cm}
\caption{Comparison of the surface profiles given by Eq.~(\ref{eq1}) [solid line] and 
Eq.~(\ref{eq3a}) [dashed line] for $A/p = 0.03$.}
\label{fig2}
\end{figure}

Here we consider the surface profile $z_0(x)$ defined in an implicit
form by
\begin{equation} \label{eq3}
z_0(x)=A\sin(qx)\exp(-qz_0(x))\,.
\end{equation}
It can also be expressed explicitly as
\begin{equation} \label{eq3a}
z_0(x) = q^{-1}\, \mathrm{W}_0 \boldsymbol( Aq\sin(qx)\boldsymbol)\,,
\end{equation}
where $\mathrm{W}_0(\zeta)$ is the principal branch of the Lambert
$\mathrm{W}$-function $\mathrm{W}(\zeta)$ \cite{lamb:58}. For small
values of $Aq$, $z_0(x)$ obtained from Eq.~(\ref{eq3a}) is very close to
the sinusoidal  grating given by Eq.~(\ref{eq1}). An example of $z_0(x)$
for  $A/p=0.03$ is shown in Fig.~\ref{fig2} together with the sinusoidal 
grating surface used in model ${\mathcal{A}}$. For both models, the
minima  and maxima of $z_0(x)$ occur at $x_{min}=(n-0.25)p$ and
$x_{max}=(n+0.25)p$,  respectively, where $n=0,\pm1,\pm2, \cdots$, but
for model ${\mathcal{B}}$  $|z_0(x_{min})| > |z_0(x_{max})|$. Since
$\mathrm{W}(\zeta)$ has a second-order branch point at $\zeta
=-\mathrm{e}^{-1}$ corresponding to $\mathrm{W}_0 = -1$ \cite{corl:96}
our considerations are limited to groove depths smaller than $A_c/p =
(2\pi \mathrm{e})^{-1} \approx 0.06$; for $A\to A_c$ we have
$z'_0(x_{min})\to\infty$.

  Then we apply the following conformal mapping:
\begin{eqnarray}
t&=&x+A\cos(qx)\exp(-qz)\,,\label{eq4}
\\u&=&z-A\sin(qx)\exp(-qz)
\label{eq5}
\end{eqnarray}
to express $\theta$ as a function of $(t,u)$. The condition $z=z_0(x)$
corresponds to $u=0$ in Eq.~(\ref{eq5}), thus,
the distortion free energy expressed in terms of the new variables
has exactly the same form as for the flat substrate
(cf. Eq.~(\ref{eq2})), i.e.,
\begin{equation} \label{eq7}
F_d[\theta]=\frac{K}{2}\int_0^p dt\,\int_0^\infty du\,
\left[(\partial_t\theta(t,u))^2+(\partial_u\theta(t,u))^2\right] .
\end{equation}
For the surface free energy we have
\begin{equation} \label{eq8}
F_s[\theta_0]=\frac{1}{2}\int_0^p dt\,
\frac{w(x)[-\sin(\theta_0(t))z'_0(x)+\cos(\theta_0(t))]^2}
{(1-qz_0(x))(1+(z'_0(x))^2)^{3/2}}\,,
\end{equation}
where $\theta_0(t)=\theta(t,u=0)$, and $x=x(t)$ is obtained from
Eqs.~(\ref{eq4}) and (\ref{eq5}) for $u=0$.

By comparing Eq.~(\ref{eq2}) with Eqs.~(\ref{eq7}) and (\ref{eq8}) we
see that due to the conformal mapping the semi-infinite system with the
rough patterned surface has been mapped onto the semi-infinite system
with a flat patterned surface and a modified form of the surface free
energy. The Green's function method can now be used to express
$\theta(t,u)$ that satisfies the Laplace equation in terms of the
boundary function $\theta_0(t)$, which allows to treat $F$ as a
functional of $\theta_0$ \cite{kond:01}. Thus, instead of solving the
Laplace equation with suitable boundary conditions (see
Eq.~(\ref{eq2a})) in the two-dimensional $xz$ plane we minimize
$F[\theta_0]$ numerically on a one-dimensional grid of variable $t$.

\subsection{Phase diagram}
\label{sec:si:pd}

\begin{figure}[t!]
\begin{center}
\includegraphics[width=8cm]{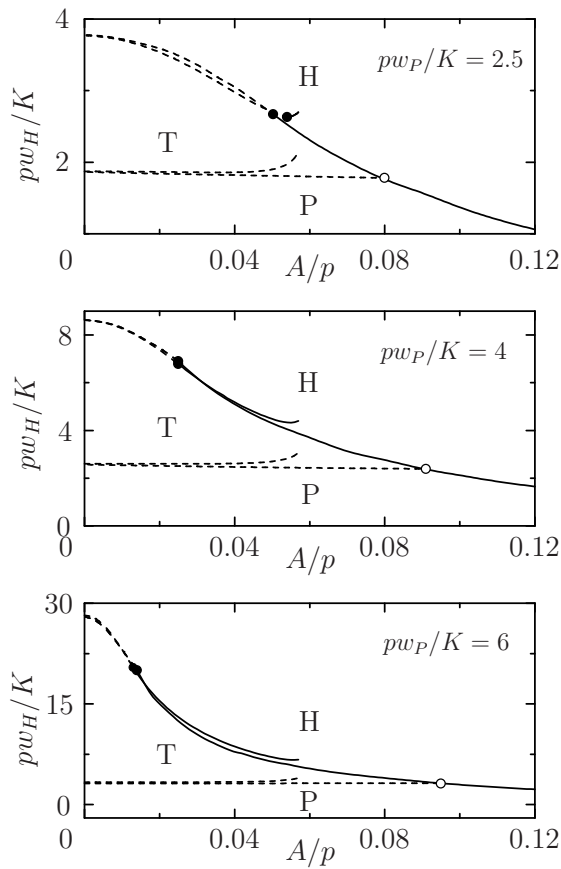}
\end{center}
\caption{Phase diagram of a NLC in contact with a single
sinusoidal grating surface [model ${\mathcal{A}}$,  see Eq.~(\ref{eq1}) and 
Fig.~\ref{fig1}] with alternating stripe pattern of locally planar and homeotropic 
anchoring as a function of groove depth $A$ and the strength of homeotropic anchoring 
$w_H$ for three values of the strength of planar anchoring $w_P$. The solid and the 
dashed lines denote first- and second-order transitions, respectively, between 
planar (P), tilted (T), and homeotropic (H) nematic textures. The solid 
and open circles mark tricritical and critical end points, respectively.
For comparison, the results obtained from model ${\mathcal{B}}$ 
[see Eq.~(\ref{eq3})] are shown with the same line code, but for $A<0.057\,p$.}
\label{fig3}
\end{figure}

Figure \ref{fig3} displays the phase diagram constructed as a function
of the groove depth $A$ and the homeotropic anchoring strength $w_H$,
for three values of the planar anchoring strength $w_P$. The textures P,
T and H are separated by the lines of phase transitions which we refer
to as the P-T, T-H and P-H transition, respectively. Both model
${\mathcal{A}}$ and ${\mathcal{B}}$ predict that the P-T transition is
always continuous, whereas the T-H transition can be either first- or
second-order, depending on the ratio $A/p$. We note, however, that the
direct first-order P-H transition, which occurs when the grooves are
sufficiently deep, follows only from model ${\mathcal{A}}$.

In the limit of small $w_H$ the planar texture is stable. When $w_H$
increases, and the grooves are shallow, the second-order P-T transition
occurs. Our calculations show that for a fixed value of $w_P$ the
location of this transition is rather independent of the groove depth,
and when $w_P$ increases it moves towards larger values of $w_H$. For
deeper grooves, the line of the continuous P-T transition terminates in
a critical end point on the first-order T-H/P-H transition line. In the
case of flat surface ($A=0$), the T-H transition is always continuous
and can exist only for $pw_P/K \lesssim 9$. If $pw_P/K \gtrsim 9$ the H
texture does not exist even for large values of $w_H$. For grating
surfaces, the T-H transition changes from the second- to first-order
when the groove depth increases. The change occurs at the tricritical
point whose position shifts to higher values  of $A$ upon decreasing
$w_P$.

We emphasize that in our model of the NLC in contact with a chemically
patterned but flat substrate ($A=0$) the phase transitions between the P
or H texture and the T texture are always continuous, and there is no
direct transition between the H and P textures. On the other hand, we
observe a first-order P-H transition for a chemically uniform sinusoidal
surface but for rather deep grooves ($A\gtrsim 0.25\, p$), similarly to
the results obtained for asymmetric surface  grating structures
\cite{brow:00}. However, as it is apparent from Fig.~\ref{fig3}, a
combination of both {\it chemical} and {\it geometrical} surface pattern
can lead to the first-order H-T transition, and also reduce the value of
$A/p$ above which the H-P transition can be found. The repercussions of
this result on possible bistable  nematic devices are discussed in Sec.
III. Note also that since the nematic texture induced by a periodic
substrate is characterized by one of the {\it effective} anchoring
directions: $\theta_a^{(eff)} = 0$, $\theta_a^{(eff)} = \pi/2$ or $0 <
\theta_a^{(eff)} < \pi/2$, the observed phase transitions can be
regarded as anchoring transitions.

In Fig.~\ref{fig3} we have also compared the predictions of models
${\mathcal{A}}$ and ${\mathcal{B}}$. We observe a good agreement for
small values of $A/p$, whereas for deeper grooves some deviations
appear. This reflects the fact that the difference between the grating
profiles defined by Eqs.~(\ref{eq1}) and (\ref{eq3}) becomes more
pronounced with increasing groove depth. Since the application of model
${\mathcal{B}}$ is limited to rather small values of $A/p$, for the
reasons already discussed, this model is unable to describe the
first-order transition between the homeotropic and planar textures.
However, the advantage of model ${\mathcal{B}}$ is that the precise
determination of phase diagrams for small ratios $A/p$ becomes feasible,
allowing one to study various aspects of the phase diagrams in more
detail and with considerably less computational effort than within model
${\mathcal{A}}$. Thus, to obtain the overall picture of the phase
transitions studied we have used information from both models.

\begin{figure}[t!]
\begin{center}
\includegraphics[width=8cm]{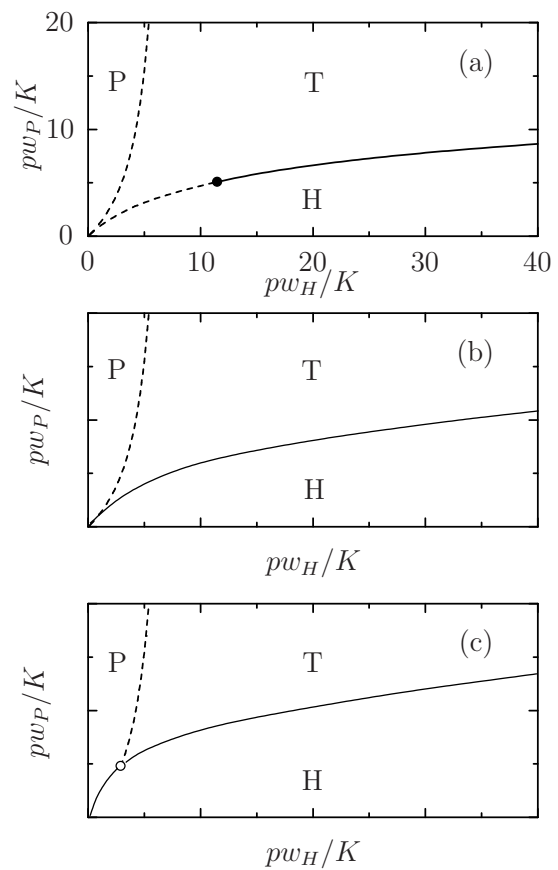}
\end{center}
\caption{Phase diagram in the $\boldsymbol(pw_H/K, \, pw_P/K \boldsymbol)$
plane for: (a) $A<A_0$, (b) $A=A_0\approx 0.075\,p$ and (c) $A>A_0$. The
meaning of symbols and lines is the same as in Fig.~\ref{fig3}. Diagram
(a) follows from the calculations based on model ${\mathcal{B}}$ for
$A/p = 0.02$. In the case of flat ($A=0$) chemically patterned substrate
(not shown here) both phase transitions are continuous everywhere (the
tricritical point on the H-T line escapes to $\infty$). Diagrams (b) and
(c) correspond to model ${\mathcal{A}}$ but they are only schematic. At
$A=A_0$ the tricritical point and the critical end point replace each
other.}
\label{fig4}
\end{figure}

\begin{figure}[t!]
\begin{center}
\includegraphics[width=8cm]{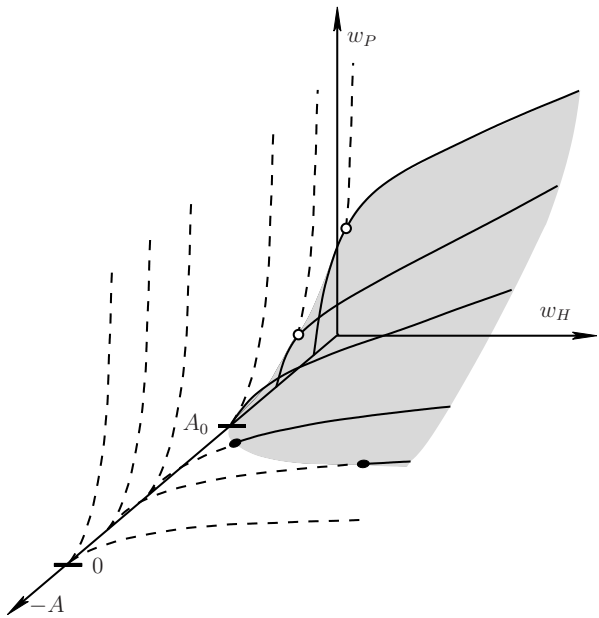}
\end{center}
\caption{Three-dimensional phase diagram in the
$\boldsymbol(pw_P/K, \, pw_H/K,\, A/p \boldsymbol)$ space. The meaning
of symbols and lines is the same as in Fig.~\ref{fig3}. The solid and
dashed lines correspond to intersections with planes $A=const$, and the
grey area corresponds to the surface of the first order transitions.
Note the evolution of the tricritical point and the critical end point
when $A/p$ changes.}
\label{fig5}
\end{figure}

In Fig.~\ref{fig4}, we present a sequence of phase diagrams in the
$\boldsymbol(pw_H/K, \, pw_P/K \boldsymbol)$ plane, for three values of
the groove depth: (a) $A<A_0$, (b) $A=A_0$ and (c) $A>A_0$, where $A_0/p
\approx 0.075 $. In all diagrams the P-T transition is continuous. In
case (a), there is a tricritical point on the H-T transition line. For
$A\to 0$ its position tends to the infinite value of $w_H$ (see
Fig.~\ref{fig5}), which is consistent with the flat substrate case. Upon
increasing $A$ the tricritical point moves towards
$\boldsymbol(pw_H/K=0, \, pw_P/K=0 \boldsymbol)$ and eventually, at
$A=A_0$ (case (b)), it reaches the origin, and the H-T transition
becomes first-order everywhere. In case (c), the continuous P-T
transition line does not extend to the origin, as in cases (a) and (b),
but terminates at the first order transition line, which for $A>A_0$
consists of two pieces: the H-T and P-H lines. Note that when $A$ passes
$A_0$ the tricritical point is replaced by the critical end point or
vice versa.

To better visualize the relation between phase diagrams presented in
Figs.~\ref{fig3} and \ref{fig4}, a schematic three-dimensional phase
diagram in the space spanned by $\boldsymbol(pw_P/K, \, pw_H/K,\, A/p
\boldsymbol)$ is also shown (see Fig.~\ref{fig5}). The two-dimensional
plots can be deduced from Fig.~\ref{fig5} by making intersections
parallel to the plane $\boldsymbol(A/p, \, pw_H/K \boldsymbol)$ or
$\boldsymbol(pw_H/K, \, pw_P/K \boldsymbol)$.

Finally we note that one can also consider the situation when
homeotropic stripes are on the sides and planar stripes are on the hills
(and in the pits) of the surface profile (see Fig.~\ref{fig1}). In such
a situation, the H-T transition line can be easily deduced from the P-T
transition line of the above-discussed case by interchanging $w_H$ with
$w_P$ and $p_H$ with $p_P$; the same holds for the line of the P-T phase
transition.

\begin{figure}[t!]
\vspace*{-1.1cm}
\begin{center}
\includegraphics[width=8cm]{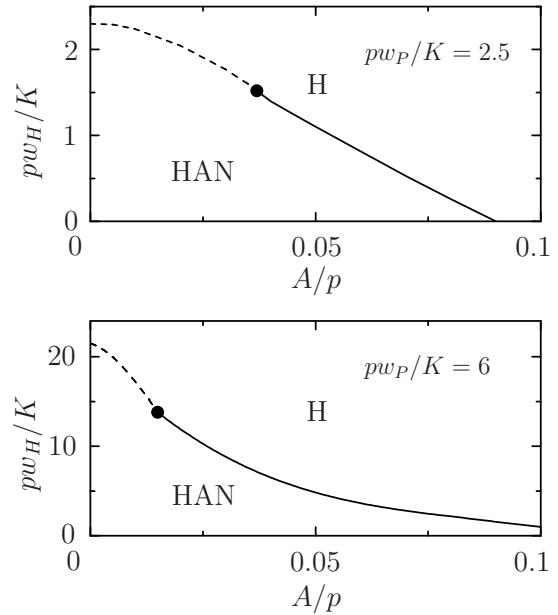}
\end{center}
\vspace*{-2cm}
\caption{Phase diagram of a NLC confined between a patterned sinusoidal
grating  surface (see Fig.~\ref{fig1}) and a flat surface with strong
homeotropic  anchoring as a function of groove depth $A$ and the
strength of homeotropic anchoring $w_H$ for two values of the strength
of planar anchoring $w_P$. The solid and dashed lines denote first- and
second-order transitions,  respectively, between a homeotropic (H) and
hybrid aligned nematic (HAN) texture.  The solid circles mark tricritical
points. The width of the cell is $D/p = 2$.}
\label{fig6}
\end{figure}

\section{Zenithally bistable nematic device}
\label{sec:finite}

We now turn our attention to a NLC confined between a chemically
patterned  sinusoidal surface (which corresponds to model
${\mathcal{A}}$) and a flat  substrate with strong homeotropic anchoring
$\theta(x,D)=0$, where $D$ is the  thickness of the cell. In the
following calculations we keep $D$ fixed, but vary  both the anchoring
strengths and the groove depth at the grating surface.  The numerical
solution of the Euler--Lagrange equation demonstrates the existence  of
two nematic director configurations, namely the homeotropic (H) texture,
where  the director field is almost uniform and perpendicular to the
flat surface, and  the hybrid aligned nematic (HAN) texture, where the
director field varies from  homeotropic to nearly planar orientation
through the cell (c.f. Fig.~\ref{fig7} below). Of course no planar
texture, which exists in the semi-infinite case, can be found  because
of the strong homeotropic anchoring on the flat substrate.

\begin{figure}[t!]
\vspace*{-0.75cm}
\begin{center}
\includegraphics[width=8cm]{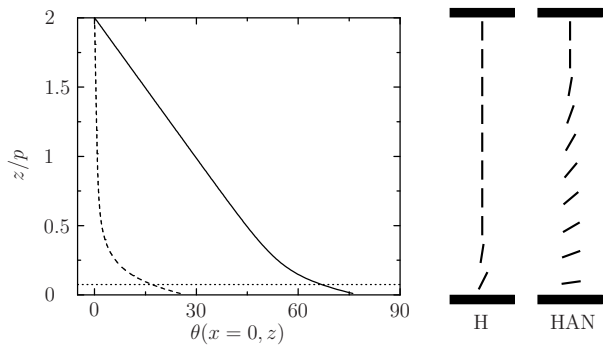}
\end{center}
\vspace*{-2.5cm}
\caption{The tilt of the nematic director $\theta(x=0,z)$ with respect
to the $z$-axis for a NLC confined between the patterned sinusoidal
grating surface shown in Fig.~\ref{fig1} and a flat surface with strong 
homeotropic anchoring. The dashed and the solid lines represent the profile 
for the coexisting H and HAN textures for the planar anchoring strength
$pw_P/K=2.45$ and groove  depth $A/p = 0.075$; see also the schematic
nematic director profiles on the right side bar. The flat surface is
located at $z = D = 2p$ while the grating surface is at $z_0(x=0)=0$ 
(see Fig.~\ref{fig1}). The location of the hills of the grating surface
is marked by the  dotted line.}
\label{fig7}
\end{figure}

Figure \ref{fig6} displays the phase diagram plotted as a function of
the groove  depth $A$ and the homeotropic anchoring strength $w_H$ on
the grating substrate  for two values of the planar anchoring strength
$w_P$ and for a fixed cell thickness  $D$. For weak homeotropic
anchoring the HAN texture is stable provided the groove  depth is
smaller than $A_{upp}$. For deeper grooves with $A>A_{upp}$ no HAN
texture  is found because distortions of the nematic director field are
too costly in the presence of the dominating homeotropic anchoring. Upon
increasing the groove depth,  the HAN-H transition changes from second-
to first-order at the tricritical point. 

The tilt angle $\theta(x,z)$ in the middle of the planar stripe ($x=0$)
is shown in  Fig.~\ref{fig7} as a function of $z$ for the coexisting H
and HAN textures. The nematic  director is almost parallel to the
$z$-axis in the H texture (for $z\gtrsim 0.6\,p$),  while it changes
smoothly from $\theta(0,z=D)=0$ to $\theta(0,z=0)<\pi/2$ in the  HAN
texture. In the latter case $\theta(0,z)$ decreases roughly linearly as
the  distance from the flat surface increases. Moreover, the numerical
calculations  exhibit that the nematic director field is rather
independent of $x$ except of a  thin surface layer (of width $\approx
p/(2\pi)$) at the grating surface. 

Finally, we emphasize that the groove depths for which zenithal
bistability is  found within the present model are considerably smaller
than those considered  theoretically ($A=0.8\,p$) \cite{parr:03,parr:04}
and used experimentally  ($A=0.35\,p$) \cite{brow:04} for a NLC confined
between a {\it pure}  geometrically patterned surface and a flat
substrate \cite{parr:03,parr:04}.  The calculated value of the average
tilt of the nematic director just above the  grating surface in the HAN
texture is comparable to the one obtained in the  case of the deeper
grooves \cite{parr:03,parr:04}.

\section{Summary}
\label{sec:summary}

We have applied the Frank--Oseen model together with the
Rapini--Papoular surface  free energy to a NLC in contact with a single
grating surface possessing an alternating  stripe pattern of locally
homeotropic and planar anchoring (Fig.~\ref{fig1}) or confined  between
the patterned grating surface and a flat substrate. Phase diagrams and
nematic  director profiles are determined numerically with the following
main results.

(1) A homeotropic, a planar, and a tilted nematic texture have been
found for the NLC  in contact with a single patterned grating surface
(Figs.~\ref{fig3} - \ref{fig5}).  Both second- and first-order
transitions between the tilted texture and the homeotropic  or planar
texture are possible. Furthermore, for appropriate values of the groove
depth  and local anchoring strengths one can also observe a first-order
transition between  the homeotropic and planar textures. It is
worthwhile to emphasize that the combination of chemical and 
geometrical surface pattern reduces the groove depth above which this
transition occurs  as compared to a pure geometrically patterned
surface. 

(2) For the NLC confined between a chemically patterned sinusoidal
grating and  a flat substrate which induces strong homeotropic
anchoring, we have determined a  first-order phase transition between a
homeotropic texture and a hybrid aligned nematic  texture for rather
shallow grooves (Fig.~\ref{fig6}). In the homeotropic texture, the 
nematic director is almost uniform and perpendicular to the flat
surface, while the  director field varies from homeotropic to nearly
planar orientation in the hybrid  aligned nematic texture
(Fig.~\ref{fig7}). Building on the results shown in  Figs.~\ref{fig6}
and \ref{fig7} it seems possible to achieve a zenithally bistable 
nematic device without the use of a deep surface grating.

It is worthwhile to emphasize that not any combination of the chemical
pattern and surface grating is capable of inducing first order
transitions between the homeotropic and tilted textures or the
homeotropic and planar textures in a semi-infinite NLC system. For
instance, if the periods of the chemical pattern $\wp$ and of the
surface grating $p$ are the same then the surface roughness merely
changes the location of the second-order transitions observed in the
flat substrate case, at least in the range of the groove depth studied
in this work. Moreover, our numerical calculations suggest that the
optimum ratio $p/\wp$ for the occurrence of the first order transitions
in the case of relatively shallow grooves is $p/\wp = 2$ shown in
Fig.~\ref{fig1}.

Finally, it is instructive to consider the situation when the chemical
surface pattern is not exactly in phase with the surface grating
profile. In such a case, for the reasons mentioned in
Sec.~\ref{sub:si:model_A}, there are no true homeotropic and planar
textures, hence the accompanied second-order phase transitions cease to
exist. However, first order transitions  between a {\it
pseudo-homeotropic} texture and the tilted or hybrid aligned texture
should still exist, where the term {\it pseudo-homeotropic} means that
the texture is characterized by $\theta_a^{(eff)}\approx 0$. Note also
that the tricritical point in the phase diagrams shown in
Figs.~\ref{fig3} - \ref{fig6} becomes a
critical point at which the difference between the pseudo-homeotropic
and titled or hybrid aligned textures disappears.

\end{document}